\documentclass{ws-procs9x6}

\begin{document}

\title{$N$* Properties from the $1/N_c$ Expansion}

\author{Richard F. Lebed}

\address{Arizona State University\\
Department of Physics and Astronomy\\ 
Tempe, AZ 85287-1504\\ 
E-mail: richard.lebed@asu.edu}


\maketitle

\abstracts{
The $1/N_c$ expansion ($N_c$ being the number of QCD colors) has been
applied in recent papers to the phenomenology of excited baryon
resonances.  This talk surveys the work done to date, and discusses
its successes and remaining challenges.}

\section{Introduction}
Baryon resonances represent a particularly striking example of
``physics in our own backyard.''  The technology exists to carry out a
vast array of interesting and incisive experiments to uncover precise
information about their nature, and researchers are limited only by
access to financial and human resources to accomplish this program.
From the theoretical point of view, however, $N^*$'s are rather
peculiar objects; indeed, for anyone approaching QCD as a pure gauge
theory of quarks confined by the gluon field, the whole hadron
spectrum is enigmatic.  One certainly expects bound quark states to
exist in some form, but fundamental (current) quarks account for only
a small portion of the hadron wave function.  Why the simple quark
model should be so successful in identifying the quantum numbers of
not only ground-state hadrons but excited ones as well is a
long-standing mystery.

In the version of this talk presented at {\it NStar 2002}, I
recognized that large $N_c$ QCD is still a rather exotic topic for
most people in our field, and raced through a 15-minute introduction
to the $1/N_c$ expansion before getting to the central issue of baryon
phenomenology.  Here I have the luxury of simply pointing to a set of
summer school lectures\cite{Prague} presented some years ago that
contain all the introductory material, and that you may peruse at your
leisure.  For the purpose of these Proceedings I reprise only the
central points:

\begin{itemize}
\item
There is nothing intrinsically crazy about letting the number $N_c$ of
QCD color charges be some number $>3$.  QCD, it turns out, would not
be qualitatively radically different were $N_c$ odd and $>3$.  Baryons
would then be fermions carrying the quantum numbers of $N_c$ quarks,
and hence would be much heavier than $q \bar q$ mesons.

\item
The $1/N_c$ expansion organizes the infinite number of Feynman
diagrams for a given process into distinct classes based on the power
of $N_c$ arising in each.  These $N_c$ factors arise from the 't~Hooft
scaling\cite{tHooft} of the strong coupling constant, $\alpha_s \!
\propto \! 1/N_c$ (which, it turns out, is the unique sensible way
to take the large $N_c$ limit), and combinatoric factors from closed
color loops.  The suppression of a class of diagrams by fewer powers
of $1/N_c$ means greater physical significance.

\item
A number of phenomenologically observed results in meson physics
follow directly from the large $N_c$ limit.  These include the
decoupling of glueballs from ordinary mesons, the OZI rule, and the
apparent dominance of heavy meson resonances over multi-pion states
({\it e.g.}, vector meson dominance), even when the latter are greatly
favored by phase space.

\item
In the large $N_c$ limit, spin and flavor symmetries for baryons
combine into a spin-flavor symmetry.\cite{GS,DM} When 3 light flavors
are included, this is the famous SU(6) symmetry.  That is to say,
SU(6) is an approximate symmetry for baryons, broken by effects of
$O(1/N_c)$.  The baryon ground states fill a multiplet that
generalizes the SU(6) {\bf 56}-plet and contains the $N$ and $\Delta$.
The $1/N_c$ expansion thus gives a field-theoretic explanation for the
successes of 1960's-vintage SU(6) results: For example, $\mu_p / \mu_n
= -3/2 + O(1/N_c)$.

\item
Baryons in $1/N_c$ may be considered in a Hartree approach, {\it
i.e.}, each quark sees (to lowest order in $1/N_c$) the collective
effect of the other $N_c\!-\!1$.\cite{Witten} Using this and the
't~Hooft scaling, it is possible to show that baryons have a
characteristic size of $O(N_c^0)$; they do not grow to arbitrarily
large dimensions as $N_c \! \to \! \infty$.

\item
It is possible to study baryon observables in the $1/N_c$ expansion by
studying operators that break the spin-flavor
symmetry.\cite{LMR,CGO,DJM2} Each such operator has a well defined
$1/N_c$ power suppression (from counting the minimum number of gluons
necessary for such an operator to appear in an interaction), and a
possible enhancement from combinatoric powers of $N_c$ if the $N_c$
quarks contribute coherently to the operator's matrix element.  Since
the number of baryons in a given spin-flavor representation is finite,
the number of operators that can give linearly independent matrix
elements, just like the basis of a vector space, is also finite.

\item
The $1/N_c$ expansion provides a natural way to define in a rigorous
way what is meant\cite{BL1} by a ``constituent quark.'' Inasmuch as
physical baryons fill well-defined spin-flavor representations whose
Young tableaux consist of $N_c$ fundamental-representation ``boxes,''
the full physical baryon wave function (as determined through
observable amplitudes) can be chopped in an unambiguous way into $N_c$
quark interpolating fields.  That is, each box represents a well
defined field whose quantum numbers are those of a quark, such that
when all $N_c$ of them are recombined, the full baryon wave function
is recovered.  Such a field may rightly be called a constituent quark;
in terms of fundamental fields it consists of many Fock components:
$q$, $q g$, $qgg$, $qq\bar q$, {\it etc.}

\end{itemize}

Using the Hartree picture and the interpretation of quark fields just
described, one may suppose that the first orbitally-excited baryons
(the ones corresponding to the negative-parity states such as
$N(1535)$, $\Lambda(1405)$, {\it etc.}) should be treated as a
spin-flavor symmetrized ``core'' of $N_c\!-\!1$ quarks and a single
quark excited to a relative orbital angular momentum $\ell=1$.  Does
this picture produce a phenomenology in agreement with experiment?
Certainly when $N_c\!=\!3$ it generates the same quantum numbers for
$N^*$'s as seen in the conventional quark model.  However, before
examining the quantitative results, let us digress briefly to see what
happens with $1/N_c$ analysis for the ground-state baryons.

The operator analysis itself is essentially a complicated version of
the Wigner-Eckart theorem.  One writes down an effective Hamiltonian
consisting of a sum over all possible linearly-independent spin-flavor
operators, including their $1/N_c$ and other [{\it e.g.}, SU(3) flavor
symmetry-breaking $\epsilon \approx 0.3$] suppressions:
\begin{equation}
{\mathcal H} = \sum_i \frac{c_i}{N_c^{n_i}} {\mathcal O}_i ,
\end{equation}
where ${\mathcal O}_i$ are spin-flavor operators whose matrix elements
are determined entirely by group theory ({\it Clebsch-Gordan
coefficients}), and $c_i$ are unknown numerical coefficients ({\it
reduced matrix elements}) that could be calculated from the dynamics
of nonperturbative QCD ({\it e.g.}, on the lattice), but can also be
extracted from experiment.

Given a set of observables, one can then determine if the $1/N_c$
expansion describes the system successfully.  Once all dimensionful
parameters are removed (for example, by taking ratios of observables),
the $c_i$'s should be of order unity.  If they are much larger, then
the $1/N_c$ expansion has failed; if much smaller, then some
undetermined physics is required beyond the $1/N_c$ expansion.  This
program was first carried out for the ground-state baryons,\cite{JL}
and the results for the isoscalar combinations are presented in
Fig.~1.  We see there that each suppression by powers of $N_c\!=\!3$
(as well as $\epsilon$) is clearly visible, consistent with the
hypothesis that the $c_i$'s are all of a ``natural'' size; one
concludes that the whole ground-state spectrum is given in a natural
way by the $1/N_c$ expansion, even for $N_c$ as small as 3.

\begin{figure}[ht]
\centerline{\epsfxsize=10cm\epsfbox{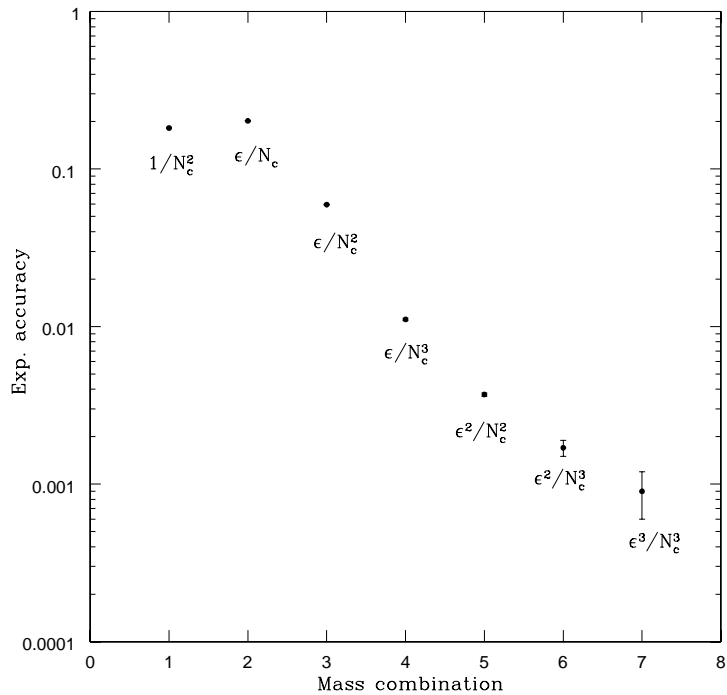}}
\caption{Isoscalar mass combinations of the ground-state baryon
multiplet in the $1/N_c$ expansion. $\epsilon \approx 0.3$ denotes
SU(3) flavor breaking.}
\end{figure}

For example, the point labeled by $\epsilon^2/N_c^2$ in Fig.~1
measures the amount by which a particular combination of
Gell-Mann--Okubo and Gell-Mann decuplet equal-spacing rules [each of
which is broken at $O(\epsilon^2)$] is violated relative to the
averaged mass of ground-state baryons (``experimental accuracy'').
The relevant operator is ${\mathcal O}_i = \{ T_8, T_8 \} /N_c$, where
$T_8$ is formed by sandwiching the Gell-Mann matrix $\lambda_8$
between the baryon quark fields.  A detailed calculation in this case
leads to the coefficient $c_i = 1.09 \pm 0.03$.  Similar results
obtain for all the other combinations.  Had we dismissed the $N_c$
factors as irrelevant, we would then have obtained $c_i \approx 1/9$
and similar power-of-3 deficits in the other mass combinations,
indicating that including the factors of $1/N_c$ is essential to
understanding the baryon mass spectrum.  It is important to note that
the old SU(6) or quark-model fits to baryon masses tended to fit each
mass individually ($p$, $n$, $\Sigma^+$, {\it etc.}), whereas this
approach fits to the smallest mass differences available, a much more
precise test of the symmetry.  Indeed, though not presented here, the
successes of $1/N_c$ continue in the isospin-breaking mass differences
as well.\cite{JL,JL2}

Many other ground-state observables, such as axial couplings, magnetic
moments, charge radii, quadrupole moments, and the spectrum of baryons
containing a heavy quark have been considered in the operator
formalism, with a high degree of success.  For sake of space, I merely
point out a recent list of references.\cite{BHL}

This conference, however, is about $N^*$'s.  To begin with, what
happens when the operator approach is applied to the $N^*$ mass
spectrum?  For much of the remainder of this talk, let us consider the
resonances in the negative-parity
multiplet.\cite{CGKM,Goity,PY,CCGL,GSS}

It is possible to carry out an operator analysis for the excited
states just as we have done for the ground states, although it is a
bit more complicated: One must distinguish operators acting upon the
$(N_c\!-\!1)$-quark core versus the excited quark and the orbital
angular momentum connecting them, and this introduces a larger
operator basis.  Nevertheless, the calculations have been done and a
remarkable result obtains: Whereas the coefficients $c_i$ for the
ground states are all $O(1)$, this is true only for a subset of the
$c_i$'s in the excited states, the remainder being much smaller.  Does
this mean that the $1/N_c$ expansion has failed here?  Not at all---in
fact, it indicates that not only are the appropriate $1/N_c$
suppressions present, but they must be enhanced by some additional
dynamical suppression (chiral symmetry, perhaps).

Table~1 demonstrates this point by presenting results of such a
fit\cite{GSS} to coefficients.  The labels $S$, $T$, and $G$ refer to
spin, flavor, and spin-flavor operators, respectively, uppercase
(lowercase) indicate those acting upon the core (excited quark), and
$\ell$ is the excited quark relative orbital angular momentum
operator.  One difference compared with the ground-state analysis is
that the $c_i$'s here have dimensions of mass and should be thought of
rather as $c_i \Lambda_{\rm QCD}$, whose natural magnitude is $\sim
500$ MeV.  The coefficients $d_i$ are those of SU(3)-breaking
operators, and should have typical sizes $\sim \epsilon c_i
\Lambda_{\rm QCD} \sim 150$ MeV.  It is clear that only
$c_{1,3,4,6,7}$ and $d_2$ appear to be of a natural size, the
remainder rather smaller.

A number of interesting conclusions follow from these results, among
which: 1)~It is perfectly natural that the $\Lambda(1405)$ is the
lightest $N^*$, despite containing a strange quark: The hyperfine
operator ${\mathcal O}_6$ does not contribute to SU(3) singlet states
but pushes all the others up 200--300 MeV.  2)~The value obtained for
the $N(1535)$-$N(1650)$ and $N(1520)$-$N(1700)$ mixing angles is
stable whether one fits the coefficients using either pionic decay,
photoproduction, or $N^*$ masses.  3)~Most significant to obtain a
good fit to mixing angles is the inclusion of the flavor-dependent
tensor [$\ell^{(2)}$] operator ${\mathcal O}_3$.  4)~The spin-orbit
coupling ($c_2$) is not large, but nevertheless explains the
$\Lambda(1520)$-$\Lambda(1405)$ splitting.


\begin{table}[ht]
\tbl{Operators ${\mathcal O}_i$ and coefficients $c_i$ (in MeV)
resulting from the best fit to the known negative-parity resonance
masses and mixings.}
{\footnotesize
\begin{tabular}{|lr@{$\, =$}r@{$\pm$}r|}
\hline
$O_1 = N_c \, {\mathbf 1} $ & $c_1$  & 449 & 2 \\
$O_2 = \ell_i \ s_i$ & $c_2$ & 52 & 15 \\
$O_3 = \frac{3}{N_c} \ \ell^{(2)}_{ij} \ g_{ia} \ G^c_{ja} $ & $c_3$ & 
116 & 44 \\
$O_4 = \frac{4}{N_c+1} \ \ell_i \ t_a \ G^c_{ia}$ & $c_4$  & 110 &  16
\\
$O_5 = \frac{1}{N_c} \ \ell_i \ S^c_i$ & $c_5$  & 74 & 30 \\
$O_6 = \frac{1}{N_c} \ S^c_i \ S^c_i$ & $c_6$  & 480 & 15 \\
$O_7 = \frac{1}{N_c} \ s_i \ S^c_i$ & $c_7$ & $-159$ & 50 \\
$O_8 = \frac{2}{N_c} \ \ell^{(2)}_{ij} s_i \ S^c_j$ & $c_8$  & 3 & 55 \\
$O_9 = \frac{3}{N_c^2} \ \ell_i \ g_{ja} \{ S^c_j ,  G^c_{ia} \} $ & 
$c_9$ &  71 & 51 \\
$O_{10} = \frac{2}{N_c^2} t_a \{ S^c_i ,  G^c_{ia} \}$ & $c_{10}$  
& $-84$ &  28 \\
$O_{11} = \frac{3}{N_c^2} \ \ell_i \ g_{ia} \{ S^c_j ,  G^c_{ja} \}$ & 
$c_{11}$ & $-44$ & 43 \\
\hline
$\bar B_1^{\vphantom{\dagger}} = t_8 - \frac{1}{2 \sqrt{3} N_c} O_1$
& $d_1$  & $-81$ & 36 \\
$\bar B_2 = T_8^c - \frac{N_c^{\vphantom{\dagger}}-1}{2 \sqrt{3} N_c }
O_1 $  & $d_2$ & $-194$ & 17 \\
$\bar B_3 = \frac{10}{N_c} \  d_{8ab}  \ g_{ia} \ G^c_{ib}  + 
\frac{5(N_c^2 -9)}{8 \sqrt{3} N_c^2 (N_c-1)} O_1$
& \multicolumn{3}{c|}{ } \\
\hspace*{2em}
$+ \frac{5}{2 \sqrt{3} (N_c-1)} O_6 + \frac{5}{6 \sqrt{3}} O_7 $
& $d_3$  & $-15$ & 30 \\
$\bar B_4 =3\ \ell_i \ g_{i8} - \frac{\sqrt{3}}{2} O_2 $ & $d_4$ &
$-27$ & 19 \\
\hline 
\end{tabular}
\label{tab1}
}
\end{table}

There have also been studies of $N^*$ production and decays using the
operator approach\cite{CC1,CC2,CC3} (A very nice review of these works
is available\cite{Carl}).

For example, one may analyze\cite{CC1} $N^* \to N \gamma$ using the
$1/N_c$ expansion, for which 19 modes have been measured.  Operators
may be classified according to the number of quark lines they connect
(In the case of Table~1, ${\mathcal O}_6$ is a 2-body operator and
${\mathcal O}_{10}$ is a 3-body operator).  Owing to the possibility
(discussed above) of coherent matrix elements it is possible, for
instance, for a 2-body operator to have the same overall power of
$N_c$ as a 1-body operator.  Such is the case for the operators
\begin{equation}
Q_* {\vec \epsilon}_* \cdot {\vec \epsilon}_\gamma \ \ {\rm and} \
\left( \sum_{\alpha \neq *} Q_\alpha \frac{\vec S_\alpha}{N_c} \right)
\cdot {\vec S}_* ({\vec \epsilon}_* \cdot {\vec \epsilon}_\gamma ) ,
\end{equation}
where * refers to the excited quark.  As before, all of the
coefficients turn out to be at most of the expected size.  However, a
detailed fit shows that the 1-body operators by themselves are
sufficient to explain the current data; the 2-body operators do not
significantly improve the $\chi^2$.  One reaches the remarkable
conclusion that the $1/N_c$ expansion again is working, but other
physics appears to be required to achieve the desired additional
suppressions of many possible terms.

Starting with this empirical observation that 1-body operators
dominate the $N^* \to N \gamma$ decays, one may now proceed to
predict\cite{CC2} quite a number (24) of $N^* \to \Delta \gamma$
amplitudes.  And while reconstructing such a process experimentally
may be a challenging task, careful analysis using the huge data set at
facilities such as Jefferson Lab can lead to the extraction of the
relevant amplitudes, and hence test the 1-body ansatz.

One may also study\cite{CC3} excited baryons in a completely symmetric
spin-flavor multiplet (what for $N_c\!=\!3$ would be called a {\bf
56}$^\prime$).  Again using the 1-body approximation, many (22)
predictions for partial widths of the processes {\bf 56}$^\prime
\to {\bf 56} +$ meson obtain.  Equally interesting are mass
predictions of the unobserved strange members of this multiplet, such
as $\Sigma^{*\prime} = 1790 \pm 192$ MeV.  One thrust of these studies
is directed toward answering the very interesting question of whether
the Roper $N(1440)$ is truly a 3-quark state ($N_c$ quarks in large
$N_c$, of course), or a mixture with hybrid $qqqg$ states, 5-quark
$qqqq\bar q$ states, or others.  A careful global analysis using mass
and decay information within the $1/N_c$ expansion may sort this out.

The conclusion one draws is that there is something special about the
$N^*$'s for arbitrary $N_c$, in that not only $1/N_c$ suppression
powers are manifest, but some other dynamics is at work minimizing the
effects of many of the possible operators.  The particular origin of
this physics is a topic currently under study.

Much has been made at this meeting about whether the quark
interactions giving rise to the baryon spectrum require flavor
dependence.  Of course, flavor exchange is a natural consequence of
meson exchange potentials, while quark potentials traditionally tend
to include spin exchange but not flavor exchange.  The $1/N_c$
approach includes both flavor-dependent and -independent operators,
and simply deduces which ones turn out to be favored (based on the
sizes of their coefficients) from fits to data.

Now, in the completely symmetric ground-state baryons (and restricted
to a fixed value of strangeness), the group theory is such that the
effect of operators with flavor dependence may always be rewritten as
arising from equivalent flavor-independent operators.  In the
mixed-symmetric negative-parity $N^*$'s, however, this is no longer
true, since the system is explicitly separated into core and excited
parts, and one may follow the flow of flavor between the two in
operators such as $\ell^{(2)} g G_c$.

But these operators have the same formal composition as the sort that
one could write down in a quark model.  For example, $\ell^{(2)} g
G_c$ represents a tensor coupling between the excited quark and the
core, where not only spin but isospin is exchanged between the two.
This can be accomplished by the excited quark trading places with a
quark in the core, a perfectly valid event in the quark model.  The
standard tensor operator in the quark model would be represented as
$\ell^{(2)} s S_c$ in this notation.  If one simply includes both
operators and lets the $\chi^2$ fit to the spectrum pick its favorite,
one finds\cite{CCGL} that the former is preferred to the latter,
meaning that flavor exchange rightfully belongs in the
phenomenological quark model for these states.

On the other hand, $\ell^{(2)} g G_c$ can occur through the exchange
of a quark-antiquark pair between the excited and core systems; a
quark moving from left to right and an antiquark moving from right to
left have the same Feynman diagram representation.  This immediately
suggests a meson exchange; however, that conclusion only holds if the
$q\bar q$ pair is correlated in a very particular way.  If the time
ordering of the two quarks is not so tightly constrained ({\it e.g.},
the $q$ is emitted by the core long before the $\bar q$), the exchange
in this single event can only be represented properly as a linear
combination involving the overlap of many meson exchanges.

So one sees that the $1/N_c$ expansion accommodates both quark and
meson pictures, and there are no contradictions between the two, if
only each picture allows for a more expansive definition of the
possible phenomena available to each.

Finally, I would like to draw your attention to brand-new
work\cite{CL1} done with Tom Cohen.  Note that the sort of analysis
used above relies on the assumption that the first band of excited
baryons consists solely of single-quark excitations of ground-state
baryons; that is, all forms of configuration mixing are assumed
suppressed.  Moreover, since real resonances are of course unstable
states with appreciable widths while the Hamiltonian used above
contains no coupling to decay channels, this analysis can strictly
only teach one about the real parts of resonant pole positions.

In fact, it is possible to study scattering partial-wave amplitudes
(wherein $N^*$'s are observed in the first place) in the context of
$1/N_c$.  It has been known for 20 years\cite{Mat} that a number of
linear relations intertwine meson-baryon scattering amplitudes at
their leading order, $O(N_c^0)$; a simple example is $S^{\pi N}_{11} =
S^{\pi N}_{31}$.  Since the $N^*$'s represent poles in these
amplitudes, the pole positions themselves must also be equal up to
$O(N_c^0)$.  That is, every $N_{1/2}$ state that couples to $\pi$-$N$
must be degenerate with a similar $\Delta_{1/2}$ state, up to
$O(1/N_c)$ corrections.

Naturally, this begs the question of whether the operator analysis of
$N^*$ masses is completely compatible with the full set of relations
among the partial-wave amplitudes.  {\it A priori\/} one might think
that our picture of $N^*$'s has been too naive, that contradictions
might arise and that would only be resolved by the inclusion of some
complicated form of configuration mixing dictated by the $1/N_c$
expansion.  But in fact the two pictures combine seamlessly\cite{CL1}
and complement each other: The amplitude relations never demand any
resonances at all, but once resonances are deemed to exist, they must
obey certain degeneracies; and the operator approach gives no
indication that there are any degeneracies at all between the given
states before the matrix elements are computed, but they nevertheless
appear and must be explained.

A remarkable result of these degeneracies is that some of the
resonances couple to certain meson-baryon channels and not others at
leading $1/N_c$ order.  For example, the state corresponding to
$N(1535)$ decays at leading order exclusively to $\eta$-$N$ rather
than $\pi$-$N$, and vice-versa for the $N(1650)$.  As experts of $N^*$
physics are well aware, the strong $\eta$-$N$ coupling for $N(1535)$
and weak one for $N(1650)$ have always been among the resonances' most
remarkable features.  Furthermore, the mixing angles between
resonances of the same $I,J$ values are predicted as simple pure
numbers at leading order ($N_c^0$).  Work in this area
continues,\cite{PS,c2} with a full treatment of $1/N_c$ corrections
next on the agenda.

\section*{Acknowledgments}
I am grateful to Jefferson Lab for travel support to the conference,
and to the organizers of the Conference for local support.  This work
was funded in part by the National Science Foundation under Grant No.\
PHY-0140362.

\end{document}